\documentstyle[11pt,newpasp,twoside,epsf]{article}
\def\etal{{\it et al.~}}
 \def\edcomment#1{\iffalse\marginpar{\raggedright\sl#1\/}\else\relax\fi}
 \reversemarginpar

\begin{document}
 \title{The Morphological and Structural Classification of Planetary 
Nebulae}
 \author{Arturo Manchado, Eva Villaver}
 \affil{Instituto de Astrof\'{\i}sica de Canarias, C/V\'{\i}a L\'actea,
 38200
      La Laguna, Tenerife, Spain}
 \author{Letizia Stanghellini}
 \affil{Space Telescope Science Institute, 3700 San Martin Drive,
 Baltimore, MD 21218, USA}
 \author{Mart\'{\i}n A. Guerrero}
 \affil{Department of Astronomy, University of Illinois, 1002 W. Green
 St., Urbana, IL 61801, USA}

 \begin{abstract}
 We present a statistical analysis
 of a complete sample (255)  of northern planetary nebulae (PNe).
Our analysis is based on morphology as a main parameter.
 The major morphological classes are: round (26\% of the sample), 
elliptical (61 \%), and bipolar (13 \%) PNe.
 About a half of the round and
30 \% of the elliptical PNe
 present multiple shells.
 Round PNe have higher galactic latitude ($|{\rm b}| =$12) and
 galactic height
 ($<{\rm z}>=$753 pc), than the elliptical ($|{\rm b}| = {\rm 7},~<{\rm
 z}>=$308 pc)
  and bipolar ($|{\rm b}| = {\rm 3},~<{\rm z}>=$179 pc). This possibly 
implies a
 different progenitor mass range across morphology,
as a different stellar population would suggest.

 \end{abstract}

 \section{Introduction}

For decades now, it has been well established that
PNe are the result of the evolution of low and intermediate mass 
stars (M$<$ 10 M$_{\odot}$). 
The actual nebular formation process has been well understood since
Kwok, Purton \& Fitzgerald (1978) explained the formation of PN as the
result of the interaction of a low density fast wind with a high
density slow wind.  
The only drawback of this {\it two wind} model is that
it can not explain well the formation of asymmetric PNe. 
Since it has been long observed
that most PNe do not have a round, regular shape, some other 
mechanism has to be invoked to produce asymmetry.
 
Mellema \& Frank (1995) implemented an interacting wind model with an
equatorial 
density enhancement. The fast low density wind interacts with an 
azimuthal dependent wind forming an asymmetrical PN.
Rotation as a way to produce asymmetric PNe has been proposed by different 
authors (e.g. Calvet \& Peimbert
 1983; Ignace, Cassinelli \& Bjorkman  1996;  Garc\'{\i}a-Segura \etal  1999).
 The presence of a magnetic field is also able to
convey an asymmetrical nebular evolution
(e.g. Pascoli 1992; Chevalier \&
 Luo 1994; Soker 1998; Garc\'{\i}a-Segura \etal  1999). 
 The common envelope evolutionary phase, typical of a close binary
star, also produces the appropriate equatorial density enhancement
to make the PN ejecta asymmetric; a similar effect can be obtained by
the presence of a substellar object in the system (e.g.
 Soker 1997).

 By correlating the
 morphological class with the different nebular and stellar
 parameters,  it may be possible to disclose
 the predominant mechanism responsible for the observed
 morphology. 

 Morphological studies of PNe have lacked a univocal
 classification scheme. Since the pioneer work by Curtis (1918),
 who discovered large structured "haloes" around some PNe, there have been
several studies all aimed at the same goal: a better understanding on the
physical meaning of nebular shapes. 
 Greig (1971) classified PNe into 15 morphological classes, ultimately 
grouped in two main classes:  binebulous and circular;
he found that binebulous PNe
 have lower galactic height distribution than circular PNe (Greig 1972).
 Zuckerman \& Aller (1986) studied a sample of 108 PNe,
 classifying them into 16 morphological types, then regrouped the
many classes into bipolar, elliptical, round, irregular, and other shapes;
 50 \% of their PN sample was bipolar, 30 \% elliptical, 15 \% round, and the
 rest
 irregular or other shapes. 
Zuckermann \& Aller (1986) could not find any correlation between the
 morphological class and the C/O abundance.
 Balick (1987) divided the morphological classes in round, elliptical,
 and butterfly. He proposed an evolutionary sequence within each 
morphological type.
Chu, Jacoby \& Arendt (1987), studied a sample of 126 extended
 PNe. They found that the frequence of multiple-shell planetary nebulae
(MSPN) was  50 \%.
 Schwarz, Corradi \& Stanghellini (1993) classified
 the Schwarz, Corradi \& Melnick (1992) sample of southern Galactic PNe
 into elliptical, bipolar, pointsymmetric, irregular, and
 stellar shapes. 
On a subsample of the same catalog of PNe, 
 Stanghellini, Corradi \& Schwarz (1993) found that the central star
 distribution was different for bipolar and elliptical PNe.
  Corradi \& Schwarz (1995), using a large PN sample ($\sim$ 400 PNe), found
 a different galactic height distribution for
 elliptical (z=320 pc) and  for bipolar (z=130 pc) PNe.

Most of these classification schemes are based on incomplete or inhomogeneous
samples. On this basis, Manchado \etal  (1996) presented a complete set of
northern Galactic PNe, to be analyzed for their morphological properties.
In $\S$2 we will illustrate this sample; we also discuss the completeness and
the morphological classification based on the sample.
$\S$ 3 presents some of the relations found between the morphology and
other nebular and stellar parameters, and includes a discussion on the 
possible evolutionary scheme for the different types of pointsymmetric PNe.
Conclusions are in $\S$ 4.
 
 \begin{figure}
 \plotone{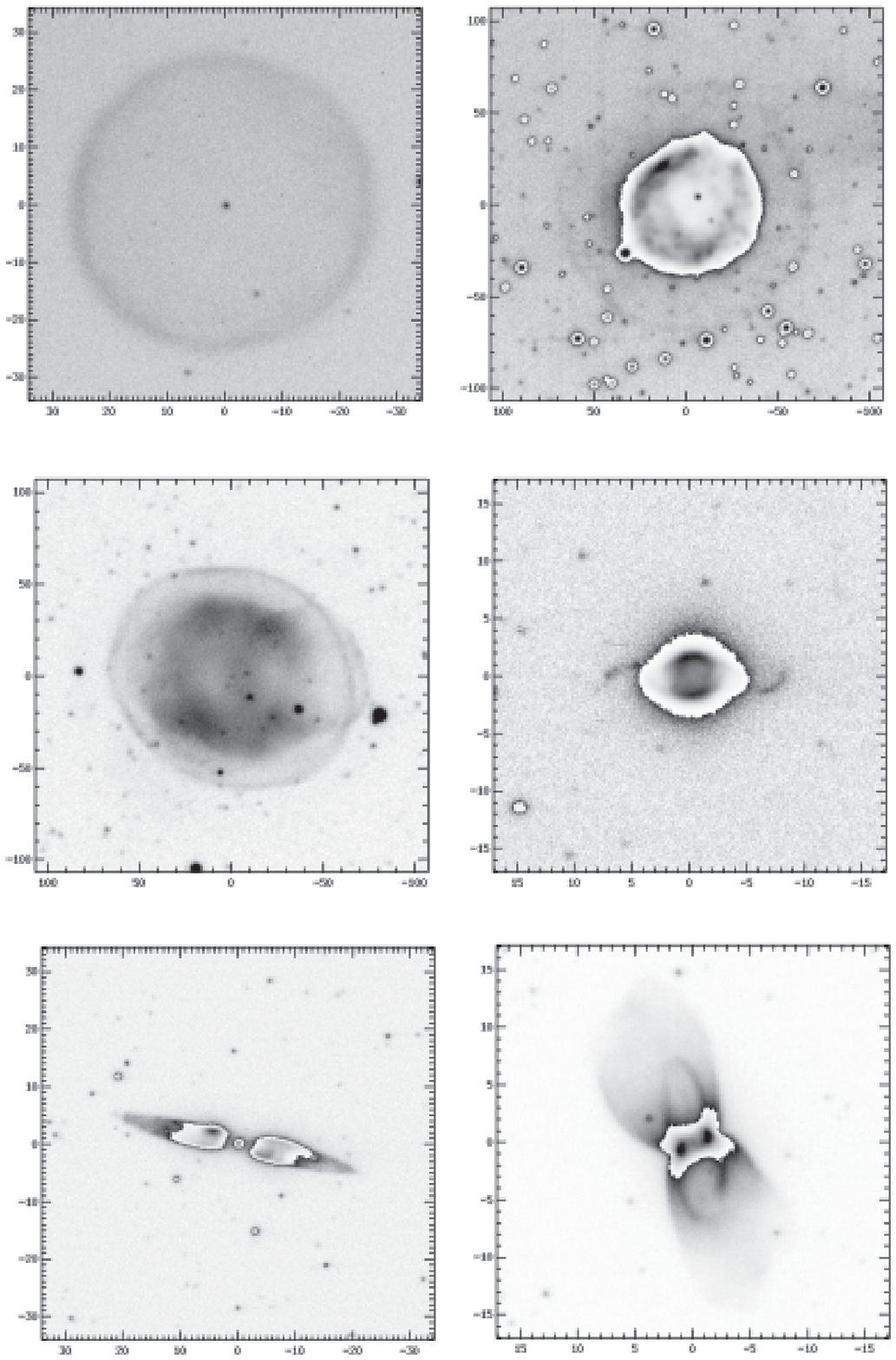}{}
 \caption{From left to right, top to bottom: narrow band images of the
PNe A 39, NGC 2438, IC 1295, He 2-429, He 2-437 and M 2-46
 }
 \label{fone}
 \end{figure}

 \section{The PN sample and its completeness}

The selection criteria for our homogeneous sample of northern 
Galactic PNe includes:
(1) all northern PNe with declination larger than --11 in the Acker \etal
(1992) catalog;
(2) all the PNe larger than 4 arc-second;
(3) images must be obtained in the narrow band filters (e.g. H$\alpha$,
[N II] or \hbox{[O III]}).
 
There are 255 PNe that fulfill these selection criteria, 205 from the
 survey by Manchado \etal (1996), 28 from Balick (1987) and 22 from
 Schwarz, Corradi \& Melnick (1992). In Figure 1 we show a selection of
these images,
representative of the various morphologies. 

 After a thorough analysis of the whole sample we decided to revise the
morphological classification by Manchado \etal (1996). In fact, to make the
individual morphological classes statistically meaningful,
we decided to make only three major morphological classes:  round (63 cases),
 elliptical
 (149 cases), and
 bipolar (43 cases) PNe. The quadrupolar PNe (7 cases) were included in the
 bipolar class, because the formation mechanisms could be very similar
(Manchado, Stanghellini \& Guerrero 1996).
  Pointsymmetry
 can be defined as a sub-class of elliptical and bipolar PNe: in fact, most 
pointsymmetric PNe have either bipolar or elliptical main shapes. A typical
case of an elliptical pointsymmetric PNe is a PN with FLIERS (e.g.
Balick \etal  1993). 
Figure 2 shows a diagram with the new morphological scheme.
 
\begin{figure}
\plotone{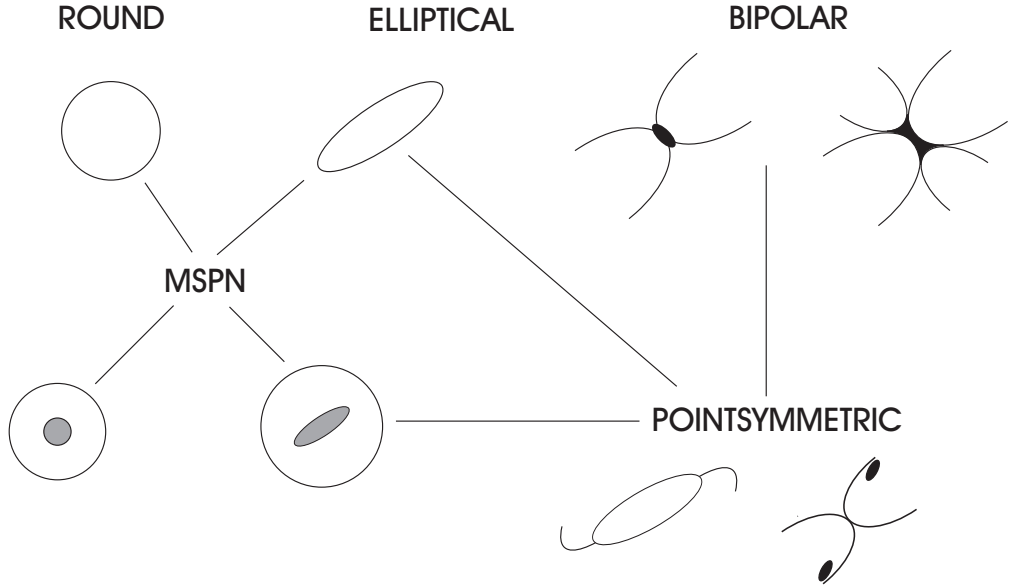}
\caption{Classification scheme
 }
\label{fone}
\end{figure}

\begin{figure}
\plotfiddle{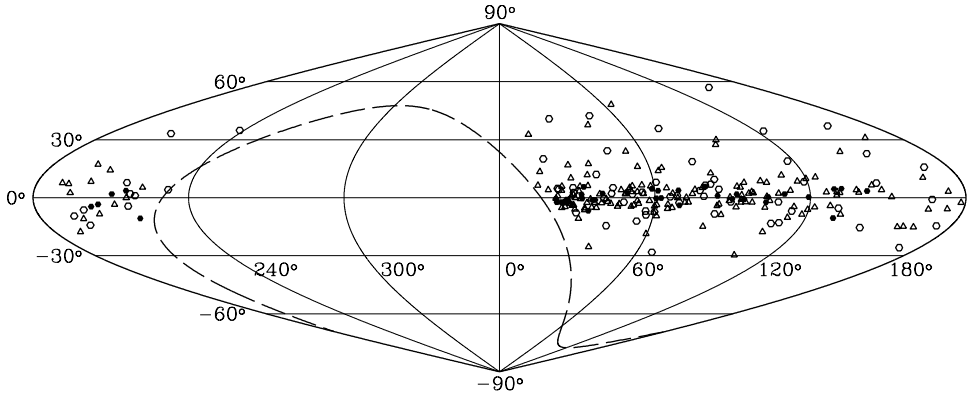}{5cm}{-90}{50}{50}{-210}{240}
\caption{Galactic distribution of elliptical (triangles), round (open circles) 
, and bipolar (filled circles) PNe.
}
\label{fone}
\end{figure}

 Although the sample is complete as far as known PNe are concerned,
 there
 may be observational biases due to a different surface brightness limit
for each
 morphological class. In order to investigate this possible bias we
 compare the statistical distribution of each morphological class
 taking
 into account the distance (Cahn, Kaler \& Stanghellini
 1992)
 and the extinction (Cahn, Kaler \& Stanghellini
 1992; Tylenda \etal 1992).
The overall distribution of morphology in our PN sample is
 58 \% elliptical, 25 \% round and 17 \% bipolar. However, we realized 
that the sample is only
 complete up to a distance of 7 Kpc for all the morphological classes. 
If we were to
 limit the statistical studies to those PNe within a distance of 7 Kpc, the 
morphological distribution 
 would be like 61 \% elliptical, 26 \% round and 13 \% bipolar. 

It can be argued that the statistical distances are not correct, so we used the
 extinction to infer completeness. 
If the sample is confined to the galactic plane, we can
 assume an extinction distance relationship of c = 0.2 per Kpc.
 Therefore, if we limit the sample to PNe with $| b| <$ 4$^{\circ}$ , 
c must be $<$ 1. In this newly defined space volume,
we find  59 \% elliptical, 28 \% round, and 13 \% bipolar PNe.
Therefore, the results
 obtained using the statistical distance and the extinction rule are very
 similar, which means that the completeness within this space volume is sound.

 \section{Relations across morphological types}

 Each morphological class was correlated with a set of nebular and stellar
 parameters from the literature (for a complete reference list, see
Manchado \etal 2000).
 
 It was found that electronic density has different values for each
 morphological class;  the median value is 1500 cm$^{-3}$ for
 elliptical, 400 cm$^{-3}$ for round, and 1000 cm$^{-3}$ for bipolar.

 Dust temperatures were derived using the IRAS 25 and 60 $\mu m$
 fluxes and dust emissivities taken from Draine \& Lee
 (1984). Both
 elliptical and round PNe have a median dust temperature of 82 K, while
 bipolar temperature is 69 K.

 The [N II]/H$\alpha$ ratio is higher for bipolar than for elliptical
 and round PNe.
 The N/O and He abundances of bipolar PNe are consistent with type I PNe
 (as defined by Peimbert \& Torres-Peimbert 1983) and in
 MSPN they are consistent with type II PNe.

 The galactic latitude distribution is different for each morphological
 class. The median of the galactic latitude is $|{\rm b}| = {\rm 7}$
 for
 elliptical, $|{\rm b}| =$12 for round and $|{\rm b}| = {\rm 3}$ for
 bipolar PNe. Figure 3 shows the galactic distribution of these three
morphological classes.  
The median values of the Galactic height are $<{\rm z}>=$308 pc for
 elliptical,
 $<{\rm z}>=$753 pc for round, and $<{\rm z}>=$179 pc for bipolar. 

By studying the pointsymmetric PNe, we find that 
for elliptical pointsymmetric the scale height is 
$<{\rm z}>=$310 pc, while the elliptical PNe without pointsymmetry have
 a $<{\rm z}>=$ 308 pc. Bipolar with pointsymmetric structure have $<{\rm
 z}>=$248 pc, while bipolar without pointsymmetry $<{\rm z}>=$110 pc.

 The different galactic height for the various morphological classes
may  imply a different stellar population. According to Miller \& Scalo
 (1979) $<{\rm z}>=$300 pc implies that the progenitor star
 has mass $<$ 1.0 M$_{\odot}$. For $<{\rm z}>=$150 pc, the mass is 
 $>$ 1.5 M$_{\odot}$, while for $<{\rm z}>=$230 pc and $<{\rm z}>=$ 110 pc
 masses will be $>$ 1.2 M$_{\odot}$ and $>$ 1.9 M$_{\odot}$.
 Therefore elliptical and bipolar PNe might have
 different distribution masses for their progenitor stars ($<$ 1.0
 M$_{\odot}$ and $>$ 1.5 M$_{\odot}$). In the bipolar
 class there is also a mass segregation. In fact, according to to their
scale height on the Galactic plane, PNe with pointsymmetric
 structure evolve from stellar masses $>$ 1.2 M$_{\odot}$, while those without
 the
 pointsymmetric structure evolve from stellar masses $>$ 1.5 M$_{\odot}$.
 These results are consistent with the other results from our 
 statistical analysis, as bipolar PNe have higher N/O
 and helium abundances.

 The fact that two different mass distributions can be inferred for bipolar
PNe, depending on the presence of pointsymmetry, can be explained with
 two evolutionary schemes for the two types:
a single, high mass star would form a bipolar PNe, due possibly to
rotation and magnetic field (e.g. Garc\'{\i}a-Segura \etal  1999),
while a bipolar pointsymmetric PN could be
due to magnetic collimation
around a precessing star (e.g. Garc\'{\i}a-Segura 1997).

 \section{Conclusions}
 A proper statistical analysis of a complete sample of PNe has allowed
 us
 to classify them into elliptical, round, and bipolar, with the
 sub-classes of
 multiple-shell and pointsymmetric PNe. It was found that 60 \% of our
PN sample present an elliptical shape, while
 26 \% are round, and 13 \% bipolar.

We use statistical distances that appear to be sound for the task. If
they are indeed correct, the different scale heights that
characterize each morphological class hint of different progenitor mass 
distribution for each class.
Two evolutionary schemes are proposed for bipolar PNe and
bipolar PNe with pointsymmetric structure.

 \section*{Acknowledgment}
 The work of EV and AM is supported by a grant of the Spanish DGES
 PB97-1435-C02-01. MAG is supported by the Spanish Ministerio de
Educaci\'on y Cultura.

 \end{document}